\documentclass[useAMS,usegraphicx,usenatbib]{mn2e}
\usepackage{graphicx}
\usepackage{amsmath}
\usepackage{amssymb}
\usepackage{subfigure}
\usepackage{url}
\usepackage{multirow}
\usepackage[linkcolor=blue,colorlinks=true,breaklinks,citecolor=blue,urlcolor=blue]{hyperref} 

\newcommand{\cm}{\rm\thinspace cm}

\newcommand{\s}{\rm\thinspace s}

\newcommand{\keV}{\rm\thinspace keV}

\newcommand{\erg}{\rm\thinspace erg}

\newcommand{\ergcmps}{\hbox{$\erg\cm\ps\,$}}

\newcommand{\ps}{\hbox{$\s^{-1}\,$}}

\newcommand{\rg}{\rm\thinspace $r_\mathrm{g}$}

\title[Comptonisation of accretion disc X-rays]{The Comptonisation of accretion disc X-ray emission: Consequences for X-ray reflection and the geometry of AGN coron\ae}
\author[D. R. Wilkins \& L.C. Gallo]{D. R. Wilkins\thanks{E-mail: drw@ap.smu.ca}\thanks{CITA National Fellow} and L. C. Gallo\\Department of Astronomy \& Physics, Saint Mary's University, Halifax, NS. B3H 3C3 Canada}
\begin{document}

\date{Accepted 2014 November 27.  Received 2014 November 26; in original form 2014 August 18}

\pagerange{\pageref{firstpage}--\pageref{lastpage}} \pubyear{2014}

\maketitle

\label{firstpage}

\begin{abstract}
We consider the Comptonisation of the photons that make up the relativistically blurred reflection that is commonly detected from the accretion discs of AGN by the coron\ae\ of energetic particles believed to give rise to the powerful X-ray continua by the inverse-Compton scattering of thermal seed photons from the disc. Recent measurements of the emissivity profiles of accretion discs as well as reverberation time lags between the primary X-ray continuum and the reflection suggest that this corona is situated at a low height above the disc and extends radially, tens of gravitational radii over the disc surface, hence should also Compton scatter the reflected X-rays. We find that the detection of blurred reflection from as close in as the innermost stable circular orbits (ISCOs) of maximally rotating black holes is consistent with such coron\ae, but requires that the corona be patchy, consisting perhaps of a number of isolated flares throughout the region. Considering only the requirement that it be possible to detect reflection from the ISCO, we find that at any given moment, the covering fraction of the inner part of the accretion disc by the corona needs to be less than 85 per cent, though allowing for the detection of `reflection-dominated' spectra in which the total reflected flux exceeds that seen in the continuum requires covering fractions as low as 50 or 25 per cent.
\end{abstract}

\begin{keywords}
accretion, accretion discs -- radiation mechanisms: non-thermal -- black hole physics -- galaxies: active -- X-rays: galaxies.
\end{keywords}

\section{Introduction}
Active galactic nuclei (AGN) are powerful extragalactic sources of X-rays. Their spectrum is dominated by a power law continuum emitted from a corona of energetic particles surrounding the central black hole. These particles are thought to be accelerated and confined by magnetic fields arising from the ionised accretion disc \citep{galeev+79,haardt+91,merloni_fabian} and they inverse-Compton scatter thermal seed photons from the accretion disc to the X-ray energies that are observed with a power law spectrum \citep{sunyaev_trumper}.

In addition to being observed directly, this coronal X-ray continuum is seen to be reflected from the optically thick, geometrically thin accretion disc \citep{george_fabian}. X-rays incident on the accretion disc are backscattered, and fluorescent lines as well as secondary emission caused by heating of the gas are produced \citep{fabian_ross_rev}. The spectrum of these reflected X-rays is blurred by relativistic effects between the emitting material in the accretion disc and the observer \citep{fabian+89}; the combination of Doppler shifts and relativistic beaming from the orbital motion of the material and the redshift from the strong gravitational field close to the black hole give emission lines (including the prominent K$\alpha$ emission line of iron at 6.4\keV) a characteristic blue shifted `horn' and extended redshifted `wing' to low energies. Measuring the extremal redshifts in the detected emission lines, the reflected X-rays are found to emerge from within a few gravitational radii (1\rg$={GM}/{c^2}$) of the black hole and identifying the inner edge of the reflection from the accretion disc with the innermost stable circular orbit enables the spin of the black hole to be measured. The supermassive black holes in many AGN are inferred to be highly spinning, with their surrounding spacetimes described by the Kerr metric with $a>0.7$) with relativistically blurred emission lines detected from the corresponding innermost stable orbit at a radius of just 1.235\rg\ \citep[\textit{e.g}][]{reynolds-13}.

The combination of reverberation time lags between variability seen in the primary continuum emission and their reflection from the accretion disc \citep{zoghbi+09,lag_spectra_paper,reverb_review} and the emissivity profile of the accretion disc, that is its pattern of illumination by the corona \citep{1h0707_emis_paper,understanding_emis_paper}, indicate an X-ray emitting corona located a few gravitational radii above the accretion disc but extending radially over the disc. Analysis of the spectrum of X-rays reflected from the accretion disc yields an emissivity profile that falls off steeply over the inner regions of the accretion disc with a power law index often exceeding 7 which then flattens over the middle region of the disc between 5 and a few tens of gravitational radii before falling off as $r^{-3}$ over the outer part of the disc. \citet{svoboda+12} consider the emissivity profile arising from the illumination of the accretion disc by a point source of X-ray emission in the corona while \citet{dauser+13} extend this treatment to a vertically collimated source extending perpendicular to the plane of the disc. Considering these two scenarios, one finds that only a compact source of X-ray emission close to the black hole can produce the observed steeply-falling emissivity profile over the inner disc. \citet{understanding_emis_paper}, however, show that to simultaneously reproduce the steep inner emissivity profile and the flattening over the middle part of the accretion disc, the corona must be extended radially at a low height over the accretion disc to around 35\rg\ in the case of the narrow line Seyfert 1 (NLS1) galaxy 1H\,0707$-$495 or to 10\rg\ in the NLS1 galaxy IRAS\,13224$-$3809 \citep{iras_fix}.

Such Comptonising coron\ae\ that cover the inner regions of the accretion disc, however, could have profound consequences for the appearance of the reflection spectrum. The reflected X-rays that are detected from the innermost stable orbit, well within the corona, will pass through the corona and will, themselves, be Compton scattered by the energetic particles found in this region. This will, in turn, change the appearance of the X-ray reflection spectrum observed from the inner accretion disc. It is therefore not immediately obvious that the detection of reflected X-rays from so close to the black hole can be self-consistently explained by the illumination of the accretion disc by such extended coron\ae.

We here discuss the effect of a Comptonising corona that extends over the inner regions of the accretion disc on the relativistically blurred X-ray reflection spectrum that arises from the disc, whether such extended coron\ae\ can self-consistently explain the X-ray spectra that are seen in AGN and the constraints that the detection of reflected X-rays from so close to the black hole places on the properties of the corona. We build upon the work of \citet{petrucci+01} who consider the effect of such a corona on the appearance of the reflection spectrum from the accretion disc, in particular the hump in the spectrum due to Compton scattering between 20 and 30\keV, to consider, in detail, the effect on the  relativistically broadened iron K$\alpha$ emission line and the modelling thereof.

\section{The Comptonisation of Photons from the Accretion Disc}
The X-ray continuum seen from accreting black holes is believed to arise from the Comptonisation of thermal photons that are emitted from the accretion disc by a plasma of extremely energetic particles, most likely electrons, around the black hole, termed the \textit{corona} \citep{sunyaev_trumper}. The seed photons are repeatedly Compton scattered as they pass through the corona and in the limit that the photon energy is less than that of the particles in the corona (characterised by the coronal temperature, $T_\mathrm{e}$), energy will be transferred from the corona to the photon field, producing an energetic X-ray continuum.

The process of Comptonisation by a thermal plasma is fully described by the Kompaneets equation \citep{kompaneets} and solutions thereof \citep{lightman+87}. The effect that the process of Comptonisation will have on the spectrum of the underlying seed photons depends on both the coronal temperature, defining how much energy is imparted to the photon field during each scattering, and the number of electrons encountered by each photon, which is conveniently measured by the optical depth to Thomson scattering, $\tau_\mathrm{e}$. The impact on the underlying spectrum of the seed photons is conveniently described by the \textit{Compton optical depth} (\textit{i.e.} the Compton $y$ parameter), which, for a non-relativistic distribution of electrons is given by
\[ y = \frac{4kT_\mathrm{e}}{m_\mathrm{e}c^2}\max\left(\tau_\mathrm{e}, \tau_\mathrm{e}^2 \right) \]
When $y \gtrsim 1$, the spectrum of the photons emerging from the corona is significantly different to that of the original seed photons. It it thus possible for Comptonisation to greatly affect the spectrum even when the electrons themselves are considered to be optically thin, for example, the Sunyaev-Zeldovich effect.

In the solution to the Kompaneets equation used by \citet{zdziarski+96} to predict the photon indices of the X-ray continua seen from accreting black holes, at photon energies much less than $kT_\mathrm{e}$, the Comptonisation of thermal photons from the accretion disc gives an equilibrium population of photons emerging with a power law spectrum that has a photon index, defined as the power law index of the photon count rate as a function of energy, $N(E)\propto E^{-\Gamma}$,
\[
\Gamma = \sqrt{\frac{9}{4} + \frac{3m_\mathrm{e}c^2}{kT_\mathrm{e}\left[\left(\tau_\mathrm{e} + \frac{3}{2}\right)^2 - \frac{9}{4}\right]}} - \frac{1}{2}
\]
While at higher photon energies, the spectrum cuts off exponentially as the coronal electrons no longer have sufficient energy to scatter a significant number of photons to higher energies. The net effect is that energy from photons more energetic than the corona is transferred back to the corona. The temperatures of AGN coron\ae\ are inferred to be in excess of 100\keV, if not 300\keV, from broad band X-ray observations made using \textit{Swift} and the \textit{Suzaku Hard X-ray Detector} \citep{burlon+2011,vasudevan+13}. In order to produce even the softest X-ray continua seen in narrow line Seyfert 1 galaxies, for example $\Gamma\sim 3$ in 1H\,0707$-$495, taking a lower bound coronal temperature $T_\mathrm{e}=100$\keV\ gives an optical depth $\tau_\mathrm{e}=0.44$ and Compton $y$ parameter $y=0.34$. A more common photon index seen among NLS1 galaxies would be $\Gamma=2$ and in this case, the harder spectrum requires $\tau_\mathrm{e}=0.97$ and $y=0.76$.

There exists a number of models for the X-ray spectral analysis package \textsc{xspec} \citep{xspec} that describe X-ray continua by the Comptonisation of seed photons. Seed photons have either a black body spectrum or a pseudo-black body spectrum where each location on the disc radiates as a black body with varying temperature according to the energy released by viscous stresses at that location, according to accretion disc models of, \textit{e.g.} \citet{shaksun} or \citet{novthorne}. One such model is \textsc{nthcomp}, which is based upon the model of \citet{zdziarski+96} and extended by \citet{zycki+99}, and solves the Kompaneets equation following the method of \citet{lightman+87}. \textsc{nthcomp} uses the \textsc{xspec} model \textsc{bbody} or \textsc{diskbb} to create the seed photon population corresponding to the black body or accretion disc pseudo-black body, respectively.

We adapted the \textsc{nthcomp} model into an \textsc{xspec} `convolution' model henceforth known as \textsc{comptonise} to model the Comptonisation of a general population of seed photons entering the corona (such as the relativistically blurred reflection from the accretion disc). Rather than using the specifically black body seed photon populations, the seed photon spectrum is taken from the combination of model components that are input to the model, but solves the Kompaneets equation for the equilibrium emergent Comptonised spectrum in the same way. It is important to note, however, that \textsc{xspec} only evaluates the input model to \textsc{comptonise} over the energy range specified by the instrument response and the Comptonisation process will scatter photons with energies originating below the lower limit of the instrument bandpass into the instrument bandpass and therefore will produce a deficit in photons at lower energies in the Comptonised spectrum. It is therefore necessary, in order to correctly reproduce the low energy part of the Comptonised spectrum to extrapolate the instrument response to lower energies. We find that extrapolating the response matrix to 50 per cent of the lowest energy of interest accurately reproduces the Comptonised spectrum at this energy. In addition, the response matrix is extrapolated upwards to 100\keV\ in order to correctly model the relativistically blurred and redshifted reflection spectrum.

\section{The Comptonisation of Reflected X-rays}
Photons that are `reflected' from the accretion disc by the processes of Compton scattering, photoelectric absorption and fluorescent line emission and bremsstrahlung that pass through an energetic corona lying above the plane of the accretion disc will be Compton up-scattered which, depending on the Compton optical depth of the corona, can have a significant impact on the spectrum of the reflected X-rays as seen by a distant observer.

Strong gravitational redshifts in the vicinity of the black hole, combined with Doppler shifts and relativistic beaming due to the orbital motion of the reflecting material in the accretion material mean that the reflection spectrum from the accretion disc seen by a distant observer is blurred, with narrow emission lines in the rest frame of the disc material being extended into a blue shifted `horn' and extended redshifted `wing' \citep{fabian+89,laor-91}. The reflected flux as a function of radius in the disc is described by the \textit{emissivity profile}. \citet{understanding_emis_paper} show that for an accretion disc illuminated by a corona at low height, extended radially parallel to the disc plane, this takes the form of a twice-broken power law, falling off steeply with index around 7 or 8 over the innermost regions of the disc, then flattening to an index close to zero at a radius of around 4\rg\ before falling off as $r^{-3}$ beyond an outer break radius that coincides with the radial extent of the corona.

The spectrum is modelled in \textsc{xspec}. The X-ray reflection spectrum itself is described by the \textsc{reflionx} model of \citet{ross_fabian} and the relativistic blurring is applied by convolving the rest frame reflection spectrum with that of a broadened emission line  using the \textsc{kdblur3} model, a modified version of \textsc{kdblur} using a twice-broken power law emissivity profile \citep{1h0707_emis_paper}.

The Compton up-scattering of the blurred reflection spectrum as it passes through the corona is described by running the spectrum through the \textsc{comptonise} model for a given coronal temperature, $T_\mathrm{e}$ and optical depth, $\tau_\mathrm{e}$ (or alternatively the temperature and photon index of the continuum that would be produced from the Comptonisation of a black body seed photon population, with the model calculating the required optical depth).

\begin{figure*}
\centering
\subfigure[$T_\mathrm{e} = 100$\keV] {
\includegraphics[width=85mm]{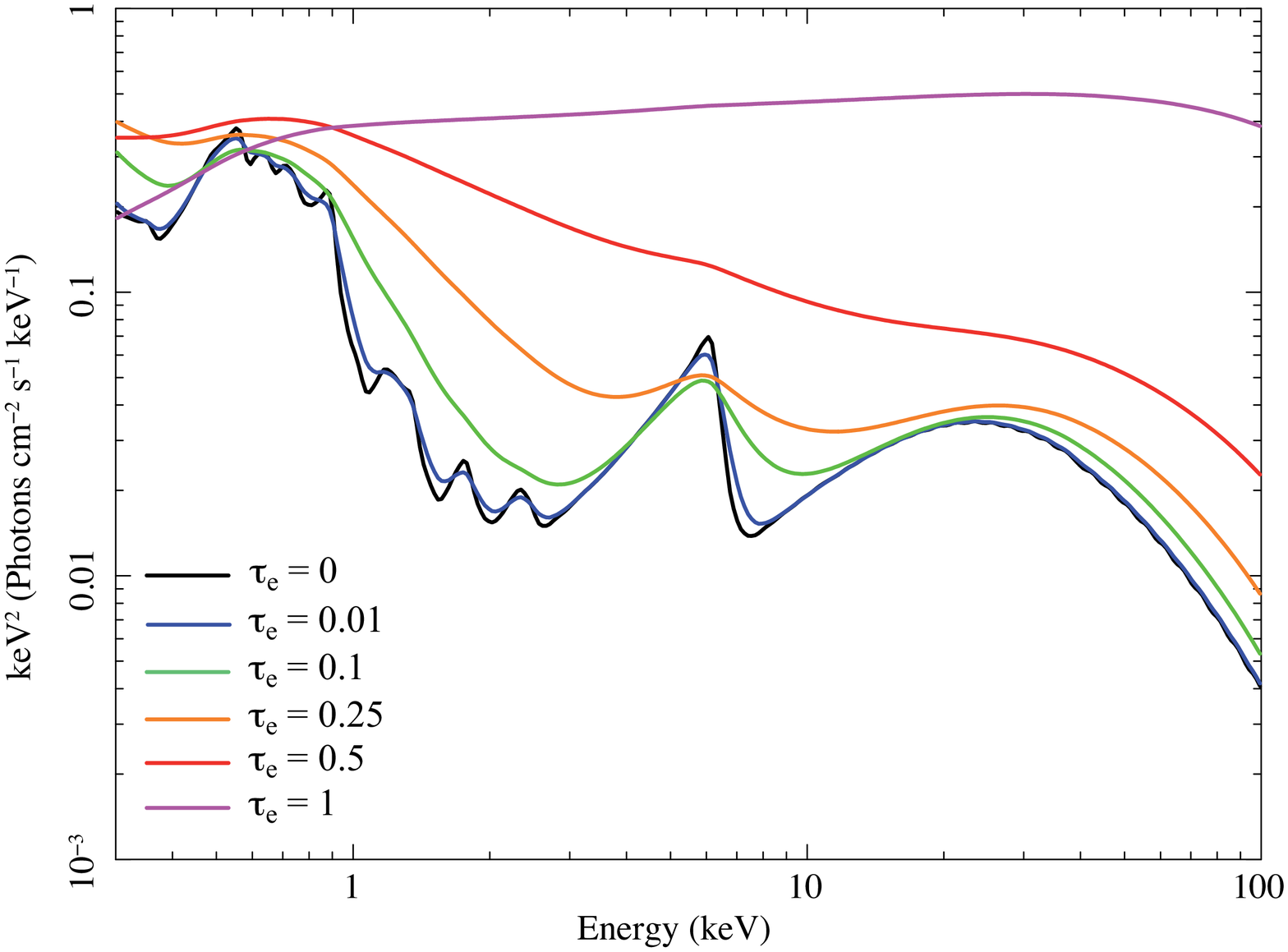}
\label{comptonised_reflection.fig:100keV}
}
\subfigure[$T_\mathrm{e} = 200$\keV] {
\includegraphics[width=85mm]{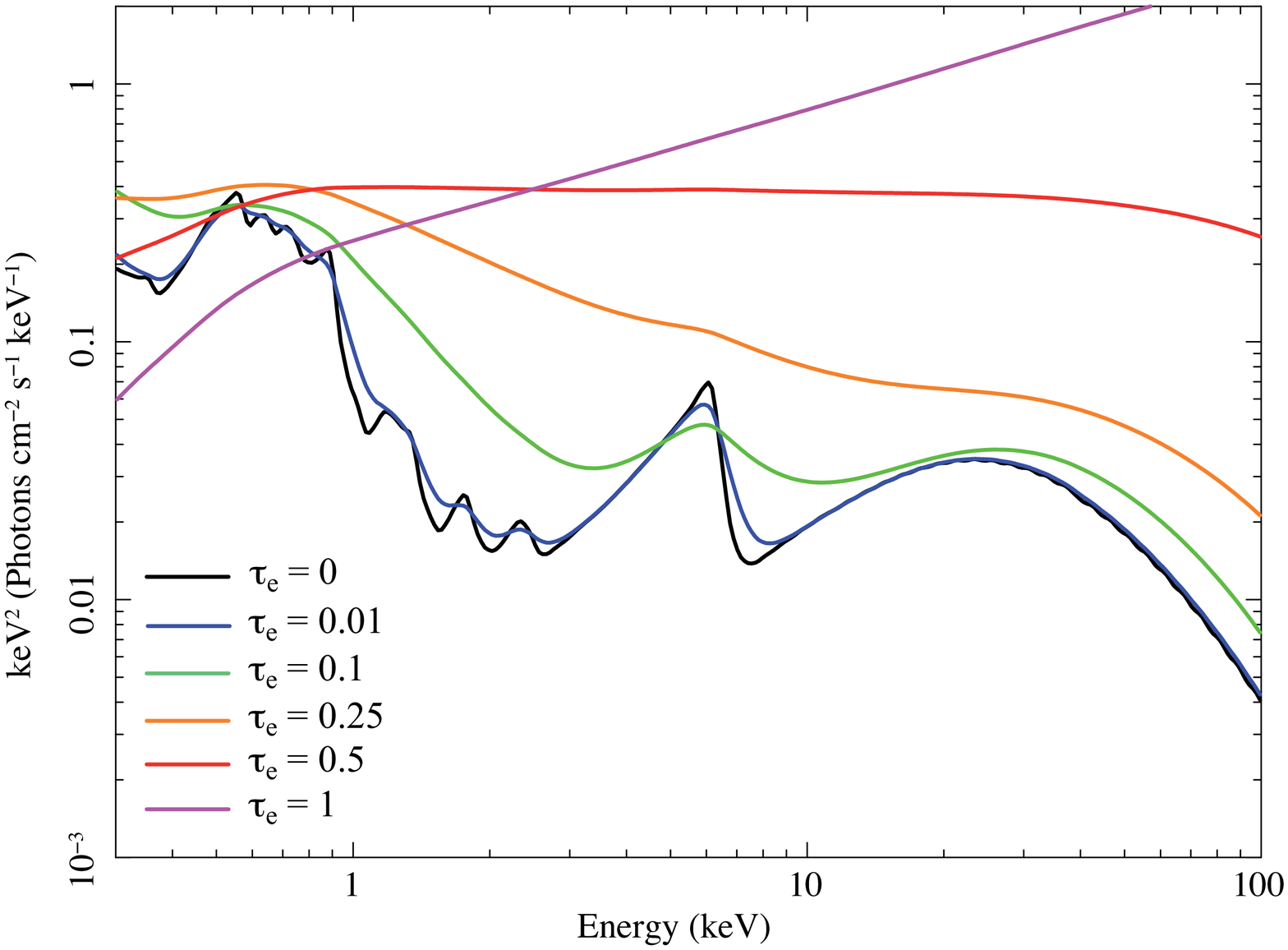}
\label{comptonised_reflection.fig:200keV}
}
\caption[]{The spectrum of relativistically blurred reflection from the accretion disc that has passed through Comptonising coron\ae\ with varying optical depths for coronal temperatures of \subref{comptonised_reflection.fig:100keV} 100\keV\ and \subref{comptonised_reflection.fig:200keV} 200\keV.}
\label{comptonised_reflection.fig}
\end{figure*}

Fig.~\ref{comptonised_reflection.fig} shows the spectrum of relativistically blurred reflection from the accretion disc that has passed through a Comptonising corona with temperature $T_\mathrm{e} = 100$\keV\ (a lower limit of that expected in the coron\ae\ of AGN) and varying optical depth to Thomson scattering $\tau_\mathrm{e}$. It is clear to see that Comptonisation has a significant effect on the spectrum of the X-rays reflected from the accretion disc. As the optical depth increases from 0 to 0.25, sharp features in the spectrum, such as the peaks of the emission lines, are smoothed out. Increasing the optical depth much further almost entirely smooths out features in the reflection spectrum and by $\tau_\mathrm{e} = 0.5$, the reflection spectrum is hardly discernible from a power law which becomes harder as the optical depth is further increased. The same is observed for higher temperature coron\ae, except that the same effects are observed at lower optical depths due to the greater energy transfer in each scattering from a hotter component (the Compton optical depth, $y$, describing how much the spectrum is altered by Compton scattering reaches the same value for a lower optical depth, for instance, we see that the spectra for $T_\mathrm{e}=100$\keV\ and $\tau_\mathrm{e}=0.5$ looks identical to that for $T_\mathrm{e}=200$\keV\ and $\tau_\mathrm{e}=0.25$ as the $y$ parameter is the same in each case).

From these results we can immediately see that in order to self-consistently produce the observed X-ray continuum by a Comptonising corona that covers at least the inner part of the accretion disc (as inferred from theoretical modelling of the measured emissivity profiles and reverberation time lags) and still be able to detect the redshifted emission from as far in as the innermost stable circular orbit and measure the spin of the black hole to be close to maximal, there must be some means by which we can observe reflected X-rays from the inner disc that have not passed through this corona. An explanation for this may be a patchy corona in which `clumps' of energetic particles are found distributed throughout the inner region though not covering the surface of the disc entirely.

\subsection{Comptonisation of the Inner Disc Reflection}

\begin{figure}
\centering
\includegraphics[width=85mm]{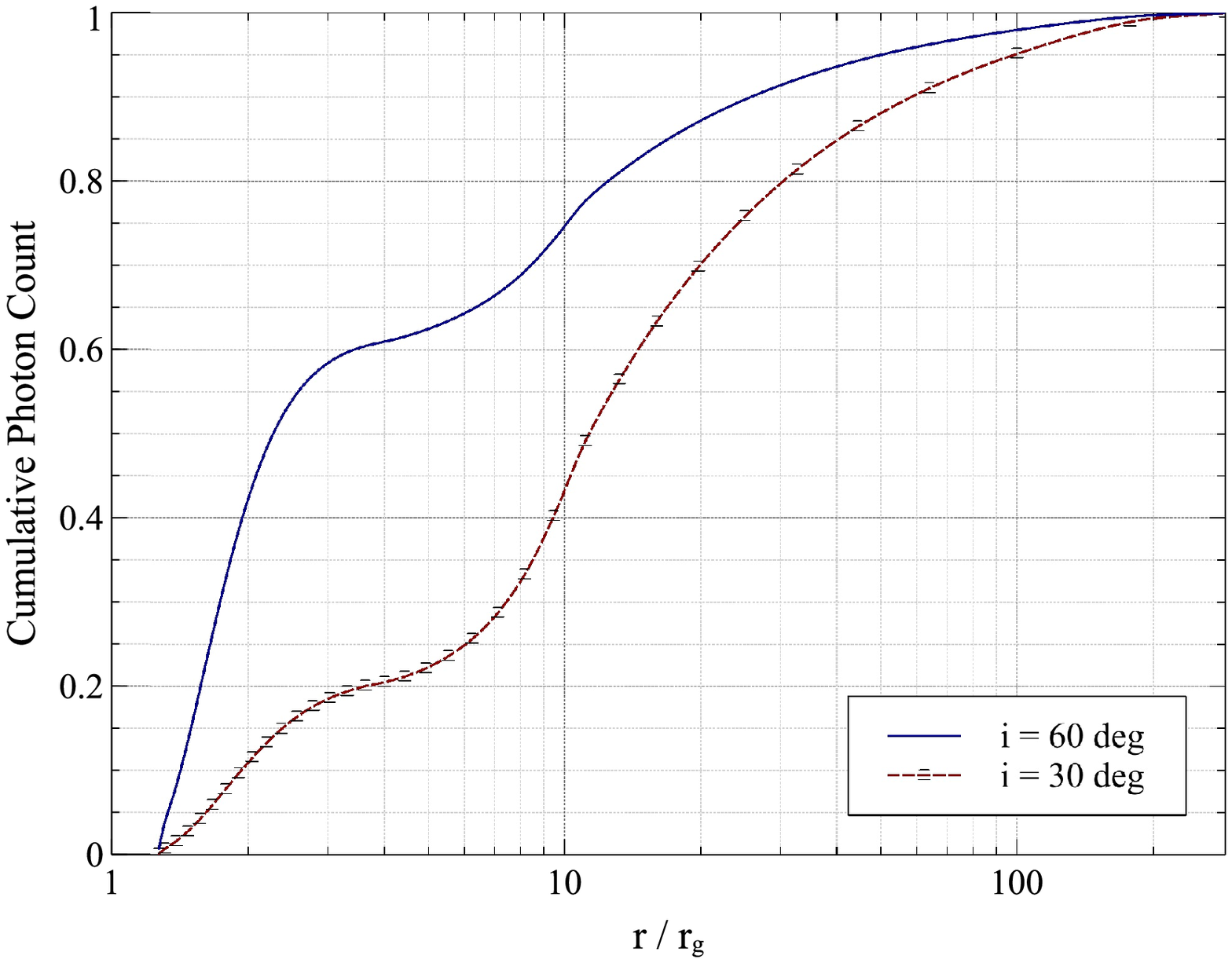}
\caption[]{The cumulative reflected flux received from an accretion disc as a function of radius, moving outward in the disc, illuminated by a corona extending radially to 10\rg\ over the surface of the accretion disc, computed by integrating the \textsc{laor} relativistic broad emission line model. Two viewing angles are considered; 30 and 60\,deg.}
\label{laor_int.fig}
\end{figure}

In order to describe the Comptonisation of the reflected photons that pass through a corona covering just the inner part of the accretion disc, it is necessary first of all to divide the blurred reflection spectrum from the accretion disc into two model components; one arising from the inner part of the disc covered by the corona that will be Comptonised, and one arising from the outer part of the disc that is observed directly. Each of these components is modelled by \textsc{cpflux}$\otimes$\textsc{kdblur3}$\otimes$\textsc{reflionx}, where each of the rest-frame reflection spectra are assumed to have the same parameters (iron abundance, incident continuum photon index and ionisation parameter) and the relativistic blurring kernels have the same inclination and emissivity profile, though use their inner and outer radius parameters to select the appropriate part of the accretion disc. The \textsc{cpflux} pseudo-model normalises the components to give the required photon count in a selected energy band and is used to split the total number of reflected photons between the components into the correct ratio, $N_\mathrm{in}$ to  $N_\mathrm{out}$, the photon arrival rates from the inner and outer parts of the disc, respectively (as if we had a single, continuous, blurred reflection component) for a given emissivity profile and inclination. Because the reflection from the disc is split into different model components, it is necessary to renormalise the flux of each to correctly match the inner and outer part of the accretion disc since the normalisation of the fundamental \textsc{kdblur3*reflionx} is hard-coded to the total flux in the entire reflection spectrum rather than per unit area of the disc. This ratio is computed from the \textsc{laor} relativistic broad emission line subroutine as it integrates the spectrum over increasing radius in the accretion disc and is illustrated in Fig.~\ref{laor_int.fig}.

The inner disc reflection component is then passed through the \textsc{comptonise} model to compute the Compton up-scattering thereof for a given coronal temperature, $T_\mathrm{e}$ and optical depth, $\tau_\mathrm{e}$, giving the full model including Comptonisation of photons reflected from the inner part of the accretion disc.
\begin{align*}
\normalfont\textsc{cpflux}^\mathrm{in}&\otimes\normalfont\textsc{comptonise}\otimes\normalfont\textsc{kdblur3}\otimes\normalfont\textsc{reflionx} \\
&+ \normalfont\textsc{cpflux}^\mathrm{out}\otimes\normalfont\textsc{kdblur3}\otimes\normalfont\textsc{reflionx}
\end{align*}
\textsc{cpflux}$^\mathrm{in}$ and \textsc{cpflux}$^\mathrm{out}$ set the ratio of the reflected photon counts arising from the inner and outer parts of the disc, respectively and it is assumed that the number of photons is conserved in the Comptonisation process (assuming the photon flux is calculated over a sufficiently large energy range, here taken to be 0.01 to 200\keV).

This model describes the total covering of the inner part of the accretion disc by the corona (though this does not necessarily mean that every photon that passes through it will undergo Compton scattering; this is determined by the optical depth which is not an unknown parameter; the optical depth of the corona is constrained to a range of values, given the temperature, to reproduce the observed photon index of the X-ray continuum). We have already ascertained, however, that if we are to detect redshifted reflection from the innermost parts of the disc, then it is necessary for some of the reflected photons to reach the observer without passing through the corona. We consider a patchy corona consisting of a number of `clumps' of energetic particles that are randomly distributed across the inner part of the disc such that the overall shape of the scattered spectrum remains unchanged (there are patches of energetic particles at all radii in an essentially axisymmetric configuration), but only a fraction $f$ of the reflected photons from the accretion disc pass through it. This is achieved by sub-dividing the $N_\mathrm{in}$ photons from the inner disc, set by \textsc{cpflux}$^\mathrm{in}$ into fractions $f$ and $(1-f)$ that are, respectively, Comptonised and seen directly. This subdivision of the photon count from the inner disc is performed by further application of the \textsc{cpflux} pseudo-model, giving the full model for the reflection component

\begin{align*}
\normalfont\textsc{cpflux}^\mathrm{in}\otimes&\left( \normalfont\textsc{cpflux}^{f}\otimes\normalfont\textsc{comptonise}\otimes\normalfont\textsc{kdblur3}\otimes\normalfont\textsc{reflionx} \right. \\
&+\left.\normalfont\textsc{cpflux}^{(1-f)}\otimes\normalfont\textsc{kdblur3}\otimes\normalfont\textsc{reflionx} \right) \\
+\ &\normalfont\textsc{cpflux}^\mathrm{out}\otimes\normalfont\textsc{kdblur3}\otimes\normalfont\textsc{reflionx}
\end{align*}

\begin{figure}
\centering
\includegraphics[width=85mm]{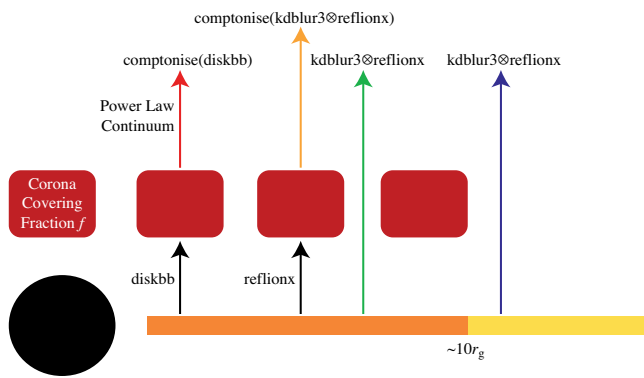}
\caption[]{Schematic of the model in which a fraction $f$ of the photons reflected from the inner region of the accretion disc pass through the patchy Comptonising corona that produces the X-ray continuum.}
\label{comptonised_reflection_scheme.fig}
\end{figure}

\begin{figure*}
\centering
\subfigure[] {
\includegraphics[height=68mm]{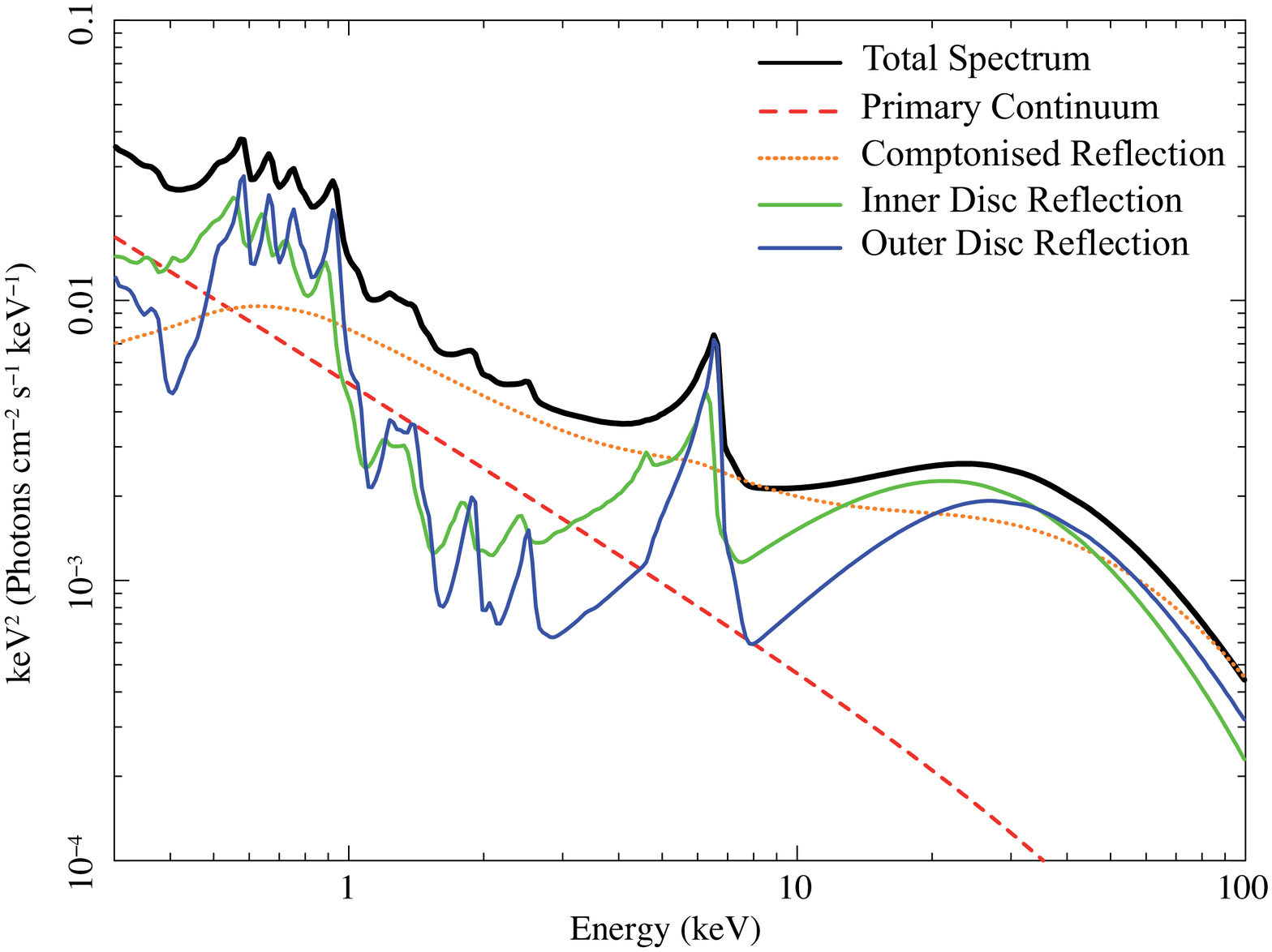}
\label{fullmodel.fig:spec}
}
\subfigure[] {
\includegraphics[height=68mm]{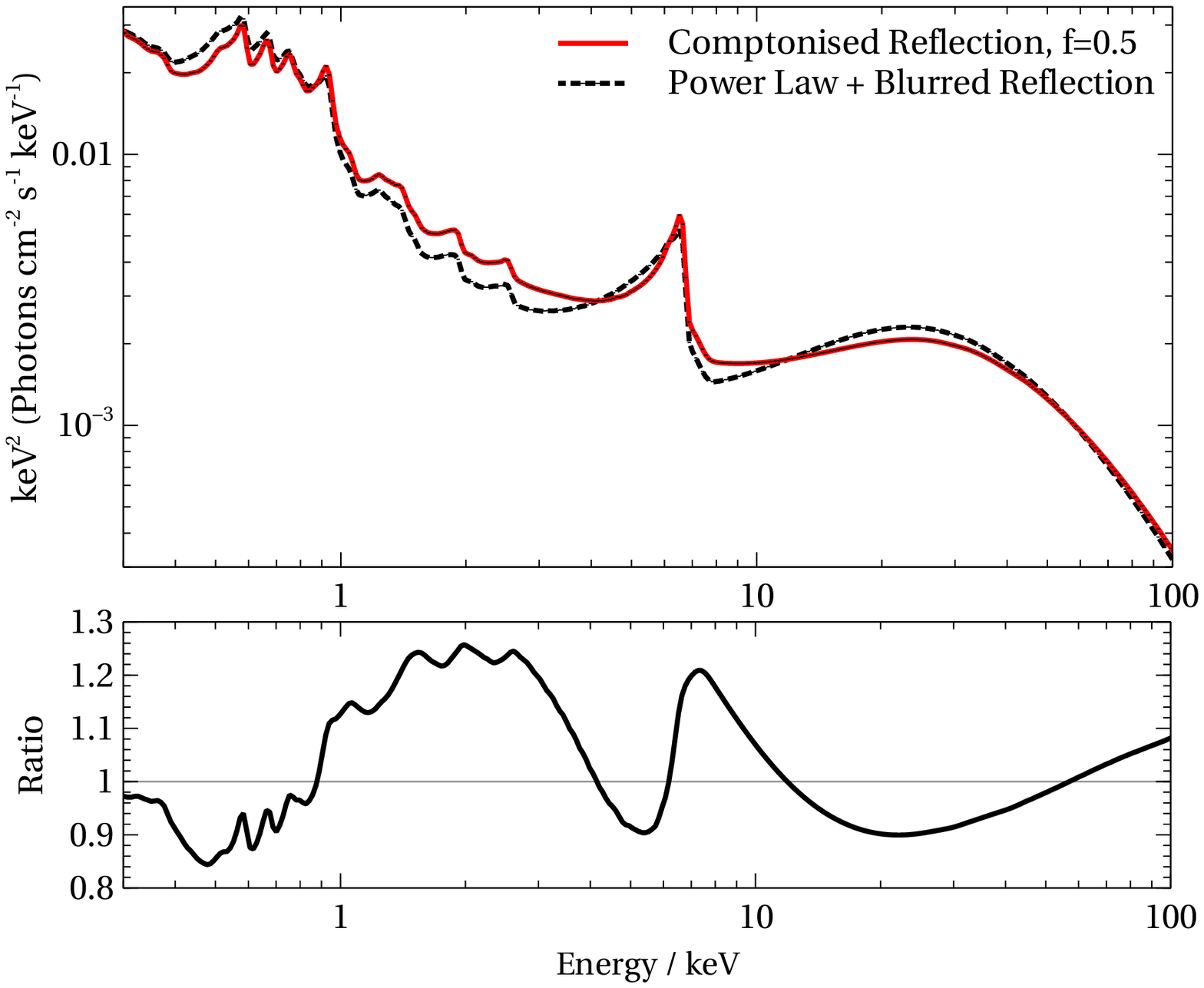}
\label{fullmodel.fig:comp}
}
\caption[]{\subref{fullmodel.fig:spec} The full spectral model consisting of an X-ray continuum from the Comptonisation of black body photons emitted from the accretion disc, the relativistically blurred reflection of this continuum from the accretion disc and the Comptonisation of 50 per cent of the reflected photons that arise from the inner region of the accretion disc that is beneath the patchy corona which extends radially out to 10\rg\ over the disc and \subref{fullmodel.fig:comp} a comparison between this model and the simple, equivalent case of a power law X-ray continuum and relativistically blurred reflection with no Comptonisation. In this example, the coronal temperature is $T_\mathrm{e}=100$\keV\ and the optical depth is set to produce a power law continuum spectrum with photon index $\Gamma = 3$ from the disc black body emission.}
\label{fullmodel.fig}
\end{figure*}

Finally, the directly observed X-ray continuum emitted from the corona is modelled by the Comptonisation of the pseudo-black body thermal spectrum emitted from the accretion disc (modelled in \textsc{xspec} by \textsc{diskbb}), and \textsc{cflux} is used to set the flux detected in the X-ray continuum and the total flux detected in the full reflection component shown above (note that in this instance we set the energy flux rather than photon count rate to more closely represent observed AGN). A schematic of the model is shown in Fig~\ref{comptonised_reflection_scheme.fig}. The full model spectrum, with the total continuum and reflected fluxes set to realistic values for an AGN as observed in 1H\,0707-495 using the model of \citet{zoghbi+09}, with ratio of the reflected to continuum flux measured over the 0.1-100\keV\ energy band, $R = F_\mathrm{refl} / F_\mathrm{cont} = 1.8$ and a fraction $f=0.5$ of the reflected photons passing through the Comptonising corona, is shown in Fig.~\ref{fullmodel.fig}.

When the reflection from the inner part of the accretion disc is Comptonised by the corona, most notably, the extended redshifted wing of the iron K$\alpha$ emission line at 6.4\keV\ is diminished. These redshifted photons arise from the innermost parts of the accretion disc, so tend to be up-scattered into a smooth Comptonised component. Looking at the ratio in Fig~\ref{fullmodel.fig:comp} between the spectrum including the Comptonisation of the reflected photons from the inner part of the accretion disc and that discounting Comptonisation, we see this causes the broad emission line to appear narrower, with a deficit in the Comptonised model around 5\keV\ and a surplus above 6.4\keV. Comptonisation also changes the shape of the redshifted wing of the line from the characteristic relativistic broad line into a smooth curve off to lower energies, with a surplus between 1 and 4\keV\ as the reflected photons, originating in both the iron K$\alpha$ line and the soft excess are Comptonised into a smooth continuum-like spectrum. Furthermore, the soft excess above the power law continuum is diminished slightly below 1\keV\ as the reflected photons contributing to this are shifted to higher energy.

Changes to the high energy part of the reflection spectrum are subtle, with a very slight shift in the Compton hump to higher energies and an overall hardening of the spectrum towards the highest X-ray energies as photons are up-scattered, though these changes to the hardest part of the reflection spectrum will likely be undetectable by real detectors.

\subsection{Consequences for Spectral Modelling}
In order to determine the importance of this effect in analysing X-ray reflection spectra in AGN, as well as to determine observational signatures that will allow the Comptonisation of the reflected photons to be measured, we simulate observations with the EPIC pn camera on board \textit{XMM-Newton}. Spectra were produced with the equivalent data quality obtained in the long observations of NLS1 galaxies \citep[\textit{e.g.}][]{zoghbi+09,iras_fix} based upon the above model, consisting of directly observed continuum emission from the Comtponisation of thermal photons from the accretion disc and its (relativistically blurred) reflection from the accretion disc. A variable fraction, $f$, of the blurred reflection from the inner part of the disc is Comptonised as it passes through the corona. Continuum and total reflected fluxes (including both the Comptonised and directly observed components) are set to realistic values we observe in AGN (in this case, taken from the model of the NLS1 galaxy 1H\,0707$-$495 of \citealt{zoghbi+09}) and the model is folded through the instrument response matrix using the \texttt{fakeit} command in \textsc{xspec}.

We then fit a simplified `standard' blurred reflection model to the simulated data consisting of only a power law continuum and the \textsc{reflionx} rest-frame disc reflection spectrum convolved with the relativistic blurring kernel \textsc{kdblur3}. This enables us to determine the impact of the partial Comptonisation of the reflected X-rays and test how well the reflection parameters can be recovered if the Comptonisation is not accounted for. The key model parameters used to simulate the observations for different fractions, $f$, of the inner disc photons that are Comptonised, are shown in Table~\ref{fit_results.tab}, along with the values recovered when simply fitting a power law continuum and blurred reflection. Two cases are considered; Comptonisation that produces an extremely steep continuum spectrum with photon index $\Gamma = 3$ as is found in the most notable sources in which these extended coron\ae\ are inferred, the NLS1 galaxies 1H\,0707$-$495 \citep{fabian+09} and IRAS\,13224$-$3809 \citep{iras_fix}, as well as a harder continuum spectrum with $\Gamma=2$, a more typical value seen across AGN and representing more extreme Comptonisation than the $\Gamma=3$ case.

It can be argued that using the parameters for 1H\,0707$-$495 and IRAS\,13224$-$3809 represent an extreme case, however the steep continuum spectrum (and, hence, lower Compton $y$ parameter) allows us to find a lower limit on the effect of Comptonisation on the appearance of the reflection spectrum while the super-Solar abundance of iron and low ionisation state of the accretion disc mean that the reflection spectrum begins with a strong iron K$\alpha$ emission line from which we can clearly judge the effects of Comptonisation. It should also be noted that it is in these extreme cases that we have the best measurements of an extended X-ray emitting corona, hence these models are the most meaningful for testing the effect of the extended corona on the reflection spectrum. 

We repeated the test with the same photon indices but with $T_\mathrm{e}=200$\keV\ (and hence a lower value of $\tau_\mathrm{e}$) and found almost identical results for each covering fraction, indicating that it is the overall effect of the Comptonisation in the corona that is important (which can be characterised by the $y$ parameter and inferred from the photon index of the continuum) rather than specifically the temperature or optical depth.

\begin{table*}
\begin{minipage}{178mm}
\centering
\caption{Key parameters of the relativistically blurred reflection spectrum that is passed through Comptonising coron\ae\ extending to a radius of 10\rg\ over the surface of the accretion disc with different covering fractions, $f$, of this region, along with the recovered values for each covering fraction when fitting a simple model consisting of only a power law continuum and relativistically blurred reflection, discounting the Comptonisation, for two different values of $\Gamma$, the photon index of the continuum produced by the corona. Parameters shown are the photon index, $\Gamma$, and the reflection fraction, $R$, defined as the ratio of the reflected to the continuum flux over the 0.1-100\keV\ band. Also, the measured inner radius of the accretion disc, the outer break radius in the twice-broken power law emissivity profile and the accretion disc inclination, $i$, as well as the goodness of fit obtained in fitting the simplified model to the spectrum, neglecting Comptonisation of the reflection. Errors are calculated within the 90 per cent confidence interval.}
\label{fit_results.tab}
\def\arraystretch{1.5}
\begin{tabular}{llcccccccc}
  	\hline
   	 & \textbf{Param.} & \textbf{Actual} & \textbf{$f=0.25$} & \textbf{$f=0.5$} & \textbf{$f=0.75$} & \textbf{$f=0.8$} & \textbf{$f=0.85$} & \textbf{$f=0.9$} &\textbf{$f=1$}\\
	\hline
	\multirow{6}{*}{Case 1} 
	
	& $\Gamma$ & 3.00 & $3.002_{-0.004}^{+0.002}$ & $2.980_{-0.004}^{+0.003}$ & $2.958_{-0.004}^{+0.002}$ & $2.960_{-0.002}^{+0.004}$ & $2.955_{-0.004}^{+0.004}$ & $2.945_{-0.003}^{+0.003}$ & $2.937_{-0.002}^{+0.002}$ \\
	
	&$R$ & 1.82 & $1.70_{-0.06}^{+0.03}$ & $1.30_{-0.05}^{+0.02}$ & $1.00_{-0.03}^{+0.03}$ & $0.95_{-0.05}^{+0.04}$  &  $0.91_{-0.04}^{+0.04}$   & $0.85_{-0.03}^{+0.03}$  & $0.77_{-0.08}^{+0.01}$  \\
	
	&$r_\mathrm{in} / r_\mathrm{g}$ & 1.235 & $1.39_{-0.07}^{+0.18}$ & $<1.35$ & $1.73_{-0.21}^{+0.05}$ & $<1.66$ & $<2.17$ & $2.21_{-0.17}^{+0.13}$ & $11.21_{-1.32}^{+1.74}$  \\
	
	&$r_\mathrm{br,out} / r_\mathrm{g}$ & 10.0 & $9.2_{-0.9}^{+0.9}$ & $8.2_{-0.2}^{+0.4}$ & $13.1_{-2.2}^{+0.6}$ & $19.1_{-1.7}^{+1.2}$ & $16.2_{-1.1}^{+1.2}$ & $19.0_{-1.6}^{+1.6}$ & $13.2_{-0.7}^{+0.6}$  \\
	
	&$i / \mathrm{deg}$ & 30.0 & $29.8_{-0.2}^{+0.2}$ & $29.4_{-0.1}^{+0.1}$ & $30.2_{-0.1}^{+0.2}$ & $30.2_{-0.1}^{+0.1}$ & $30.2_{-0.1}^{+0.2}$ & $30.1_{-0.1}^{+0.3}$ & $29.7_{-0.1}^{+0.1}$  \\
	\cline{2-10}
	&$\chi^2 / N_\mathrm{DoF}$ & --- & 1.10 & 1.12 & 1.21 & 1.23 & 1.27 & 1.28 & 1.27 \\
	\hline
	\multirow{6}{*}{Case 2} 
	
	& $\Gamma$ & 2.00 & $2.006_{-0.002}^{+0.002}$ & $1.967_{-0.003}^{+0.003}$ & $1.968_{-0.003}^{+0.003}$ & $1.965_{-0.003}^{+0.003}$ & $1.949_{-0.003}^{+0.004}$ & $1.949_{-0.004}^{+0.003}$ & $1.943_{-0.002}^{+0.002}$ \\
	
	&$R$ & 1.82 & $1.66_{-0.02}^{+0.04}$ & $1.20_{-0.02}^{+0.02}$ & $0.91_{-0.01}^{+0.01}$ & $0.89_{-0.02}^{+0.02}$  &  $0.84_{-0.02}^{+0.02}$   & $0.78_{-0.04}^{+0.007}$  & $0.71_{-0.007}^{+0.007}$  \\
	
	&$r_\mathrm{in} / r_\mathrm{g}$ & 1.235 & $1.57_{-0.08}^{+0.08}$ & $1.61_{-0.05}^{+0.05}$ & $<1.32$ & $<1.26$ & $3.37_{-0.63}^{+0.17}$ & $3.28_{-0.50}^{+0.35}$ & $9.27_{-0.98}^{+1.10}$  \\
	
	&$r_\mathrm{br,out} / r_\mathrm{g}$ & 10.0 & $9.3_{-0.6}^{+0.5}$ & $10.5_{-0.8}^{+0.9}$ & $11.8_{-0.2}^{+0.7}$ & $11.9_{-0.2}^{+0.9}$ & $12.8_{-0.5}^{+0.5}$ & $13.5_{-0.6}^{+0.5}$ & $<14.7$  \\
	
	&$i / \mathrm{deg}$ & 30.0 & $30.0_{-0.1}^{+0.1}$ & $30.1_{-0.2}^{+0.2}$ & $30.1_{-0.1}^{+0.1}$ & $30.2_{-0.2}^{+0.2}$ & $30.2_{-0.2}^{+0.2}$ & $30.2_{-0.2}^{+0.1}$ & $29.9_{-0.2}^{+0.2}$  \\
	\cline{2-10}
	&$\chi^2 / N_\mathrm{DoF}$ & --- & 1.25 & 1.07 & 1.15 & 1.06 & 1.15 & 1.28 & 1.28 \\

	\hline
\end{tabular}
\end{minipage}
\end{table*}

The results of fitting the simplified reflection model to simulated spectra in which the reflection from the accretion disc is partially Comptonised, show that while, in general, an acceptable fit is found, as the fraction of the relativistically blurred reflection that is Comptonised increases, the fit to the spectrum by this simplified model becomes progressively worse. Despite this worsening fit, however, the parameters of the reflection spectrum are, in general, still recovered successfully from the fit (\textit{i.e.} the detected shape of the remaining reflection spectrum is not greatly altered).

As long as the covering fraction of the inner accretion disc by the patchy corona is below around 80 per cent, significant reflection from the innermost regions of the accretion disc is still detected and the inner radius of the accretion disc that is measured is close to the actual value for, in this case, a maximally rotating Kerr black hole, with up to around a 0.5\rg\ error. So long as the covering fraction remains below around 80 per cent, the spin of the maximally spinning black hole will still be successfully measured to $a>0.95$. Once $f\gtrsim 0.85$, the significant Comptonisation of the reflection that does arise from the inner disc means that the relativistically blurred reflection features become less detectable. The measured value of the inner disc radius increases until, at the highest covering fractions, no reflection is detected from the inner disc and the inner radius is measured to be the edge of the corona. This enables the model spectrum to reproduce the remaining broad line that is detected arising from outside this radius.

\begin{figure*}
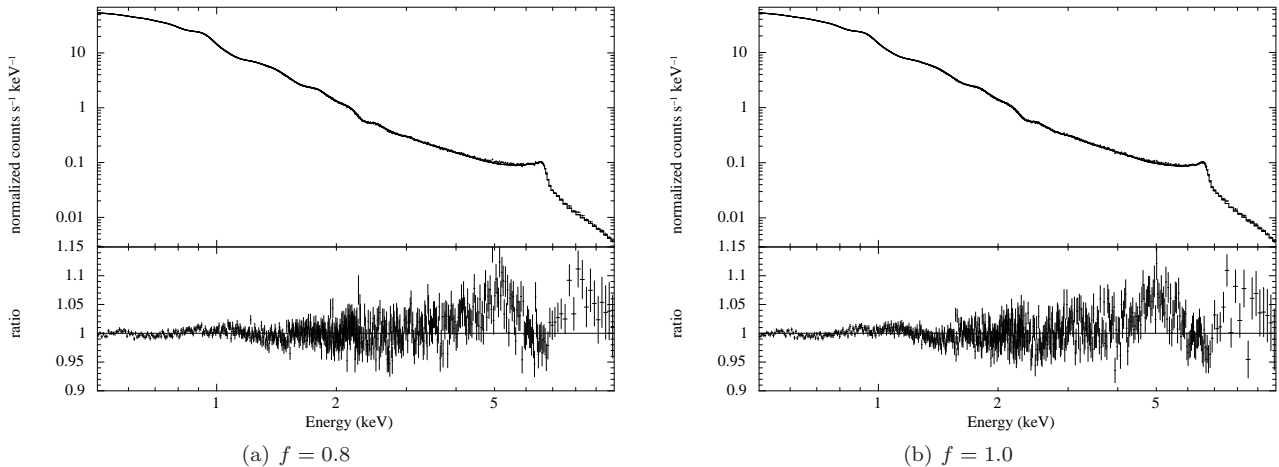

\centering
\subfigure[$f=0.8$] {
\includegraphics[height=85mm,angle=270,trim=0 0 -5mm 0]{spec_ratio_80.ps}
\label{ratios.fig:80}
}
\subfigure[$f=1.0$] {
\includegraphics[height=85mm,angle=270,trim=0 0 -5mm 0]{spec_ratio_100.ps}
\label{ratios.fig:100}
}
\caption[]{Simulated spectra, including for the Comptonisation of a fraction $f$ of the photons arising from the inner accretion disc ($r<10$\rg), for the cases \subref{ratios.fig:80} $f=0.8$ and \subref{ratios.fig:100} $f=1.0$. The spectra are fit with the simplified model consisting of just a power law continuum and relativistically blurred reflection from the disc with no Comptonisation thereof, to demonstrate the residuals that might be expected. Most notably, the core of the iron K$\alpha$ line between 4 and 6\keV\ is underestimated by the simple model if Comptonisation of the reflected X-rays is significant.}
\label{ratios.fig}
\end{figure*}

It is still possible to detect the reflection that does escape without passing through the corona at such high covering fractions because the coron\ae\ required to produce the hard X-ray continua observed in AGN (with $\Gamma < 3$) have a sufficient Compton optical depth that the spectrum of the reflected photons that are scattered is reduced to an almost featureless power law (slightly harder than the continuum spectrum, since the reflected continuum photons are up-scattered further). So long as a detectable number of reflected photons can escape from the inner disc without being scattered, the spectral features that constrain the inner radius of the disc will still be detectable, albeit diminished with respect to the peaks of the broad emission lines from regions of the disc outside the corona. This causes the worsening fit at higher coronal covering fractions, as the model fits to the diminished red wing of the line, leaving residuals around 5\keV\ (Fig.~\ref{ratios.fig}) approximately resembling a Gaussian line but at too low energy to attribute to unblurred reflection from distant material. The smooth spectrum of the Comptonised reflection means it will be detected using the simple reflection model as part of the continuum, explaining the very slight hardening seen in the spectrum as the covering fraction is increased.

The clearest manifestation of the Comptonisation of photons that are reflected from the inner regions of the accretion disc is the underestimate in the measured reflection fraction, $R$. It should be noted that in the simulated spectra, the reflected flux used to compute the reflection fraction is defined to include all of the primary continuum photons that are incident on the accretion disc and hence includes the Comptonised part of the reflection.

Given a measurement of such low reflection fractions, Fig.~\ref{ratios.fig} shows the residuals to the simple blurred reflection model that might be expected if there is significant Comptonisation of the reflected X-rays that is not accounted for. The simplified reflection model does not fully account for the shape of the iron K$\alpha$ line and to obtain the best statistical fit to the data, reproduces the redshifted wing of the emission line. This wing is diminished by the Comptonisation of a portion of the photons reflected from the inner disc. Because the model is fitting the reduced redshifted wing of the line ($3-4$\keV), it consequently underestimates the reflection appearing in the in the core of the line between 4 and 6\keV. The core of the line is predominantly emitted from the outer regions of the accretion disc and is, thus, not diminished by Comptonisation. We find that the core of the line is underestimated even when the Comptonisation is not total such that the inner radius of the accretion disc is not overestimated ($f=0.8$). When the Comptonisation is so severe that the accretion disc appears truncated, this residual is still apparent (as well as a very subtle residual at the extremal redshift of the line at the inner edge of the disc) although less pronounced given the non-detection of the reflection from the inner part of the disc in this case, so the model is better matched to the outer disc reflection. The partial Comptonisation of photons reflected from the inner part of the accretion disc also produces excess emission between 7 and 10\keV\ as more photons are shifted to higher energy.

We also note that with increased fractions of the inner disc reflection that are Comptonised, there is a slight overestimate (up to around 5\rg) of the radius of the corona from the emissivity profile measured via the relativistically broadened emission lines. This is likely occurring due to the decreased flux seen from the inner disc compared to the outer disc in the line and overestimating the outer break radius in the twice-broken power law emissivity profile goes some way towards compensating for this, increasing the emission from the outer disc relative to the inner. This is observed particularly in the case in which $\Gamma=3$ where the break radius oscillates between $f=0.75$ and $f=0.85$ rather than changing smoothly. This oscillation is less than the statistical error suggesting that the Comptonisation of the reflected photons is introducing a systematic error in this parameter due to the way in which the profile of the broad line is altered and is not adequately modelled by simple relativistically blurred reflection. A similar, though smaller systematic error is evident in the inner radius of the accretion disc with the upper limit on this parameter for $f=0.8$ less than that for $f=0.75$.

This overestimate in the break radius, however, is likely not significant in the measurements of the corona made in 1H\,0707$-$495 and IRAS\,13224-3809 since the high reflection fractions measured in these sources indicate that the covering fraction of the inner disc by the corona is not sufficiently high to cause this. Measurements of other parameters related to the relativistically blurred reflection are, however, found to be unaffected by the partial Comptonisation of the inner disc reflection, with the iron abundance (8 times the Solar abundance), ionisation parameter (50\ergcmps), inclination ($30\deg$) and the remaining parameters of the broken power law emissivity profile (the three power law indices of 7, 0 and 3 for the inner, middle and outer sections of the emissivity profile, and inner break radius at 4\rg) recovered well in each case.

\section{Implications for the Corona}
Relativistically blurred reflection is detected from the inner regions of the accretion discs around a vast number of black holes in both AGN and Galactic X-ray binaries. Moreover, in many of these objects, the inner disc radius is constrained, through the X-ray spectrum, to be close to the innermost stable circular orbit for black holes close to maximal spin. In a growing number of these objects with a relatively unobscured view of the accretion disc by absorption along the line of sight, high quality X-ray spectra have allowed the emissivity profile of the accretion disc to be measured, which, when combined with measurements of reverberation time lags between variability in the X-ray continuum and its reflection from the disc, imply X-ray emitting coron\ae\ at a relatively low height above the accretion disc but extending of order 10\rg\ over the disc. Not only that, but if a substantial X-ray continuum is to be produced from the Comptonisation of thermal seed photons arising from the accretion disc, then the corona must have a significant cross section over the plane of the disc in order to scatter a significant number of seed photons.

If such coron\ae\ are to produce the observed X-ray continua by Comptonisation and we are to detect the relativistically blurred reflection spectra that are observed in these sources, extending down to the innermost stable orbit, well within the corona, we infer that spatially extended corona must be patchy. That is to say that there are clumps of energetic particles essentially randomly distributed throughout the `corona' and a certain fraction of the reflected photons from the disc are able to escape without being scattered from all radii on the disc. Taking only the requirement that it is possible to measure an inner disc radius close to the innermost stable circular orbit of a maximally rotating black hole, well within the corona, we find that the fraction of the inner region of the accretion disc at any given moment need only be less than around 85 per cent, so there can still be a significant part of the inner disc covered and, thus, there is no problem producing a luminous X-ray continuum since there is still a significant cross-section for interaction with the thermal seed photons from the accretion disc.

A more stringent constant, however, can be derived from the reflection fraction. It is difficult to know \textit{a priori} what the reflection fraction should be as this will depend upon the geometry of the corona. If the corona is compact and close to the black hole, then many of the continuum photons emitted will be focussed towards the black hole and hence on to the inner regions of the disc rather than being able to escape to contribute to the directly observed continuum and the reflection fraction can be as high as $R\sim10$ \citep[\textit{e.g.}][]{1h0707_jan11}. In this case, however, there cannot be significant Comptonisation of the reflected photons as the corona covers, at most, the inner 2\rg\ of the disc. For coron\ae\ extending to 10\rg\ over the disc, general relativistic ray tracing simulations using the code of \citet{understanding_emis_paper} that count the number of photons that are able to escape to be observed in the continuum compared to the number that hit the accretion disc, predict a reflection fraction around 2. Assuming the measured value of 1.8 for the NLS1 galaxy 1H\,0707$-$495 with the spectral model of \citet{zoghbi+09} is a 10 per cent underestimate of this value due to the Comptonisation of the reflected photons, we can limit the covering fraction of the corona to between 25 and 50 per cent (we constrain the covering fraction only to lie within this broad range as the exact value will depend upon the exact extent geometry of the corona as well as the inclination of the accretion disc to the line of sight, which determines the number of reflected photons that are observed from the inner part of the accretion disc, while a broad range is illustrative of the coron\ae\ of a number of NLS1 galaxies). Thus, the corona can be considered significantly `patchy' though with a non-negligible covering fraction such that it remains the plausible origin of the prominent X-ray continua that are observed.

When a large fraction of the reflection from the inner disc is Comptonised, very low reflection fractions (less than unity) can be measured. Theoretically, in the absence of gravitational light bending, any corona (whether point-like or extended) above an accretion disc extending to infinity (hence subtending a solid angle of $2\pi$ to an observer within the corona) will never be less than approximately unity, thus measuring such low reflection fractions requires some explanation. One possible explanation is that the primary X-ray source is moving at relativistic speeds away from the disc (\textit{e.g.} the base of a jet) which would cause the primary X-ray emission to be beamed away from the disc \citep[\textit{e.g.}][]{beloborodov} and towards the observer, reducing the measured reflection fraction. Or, we see here, that low reflection fractions can be explained by a corona that covers a sufficient fraction of the inner disc to Comptonise the reflection that arises from this region.

\begin{figure}
\centering
\includegraphics[width=85mm]{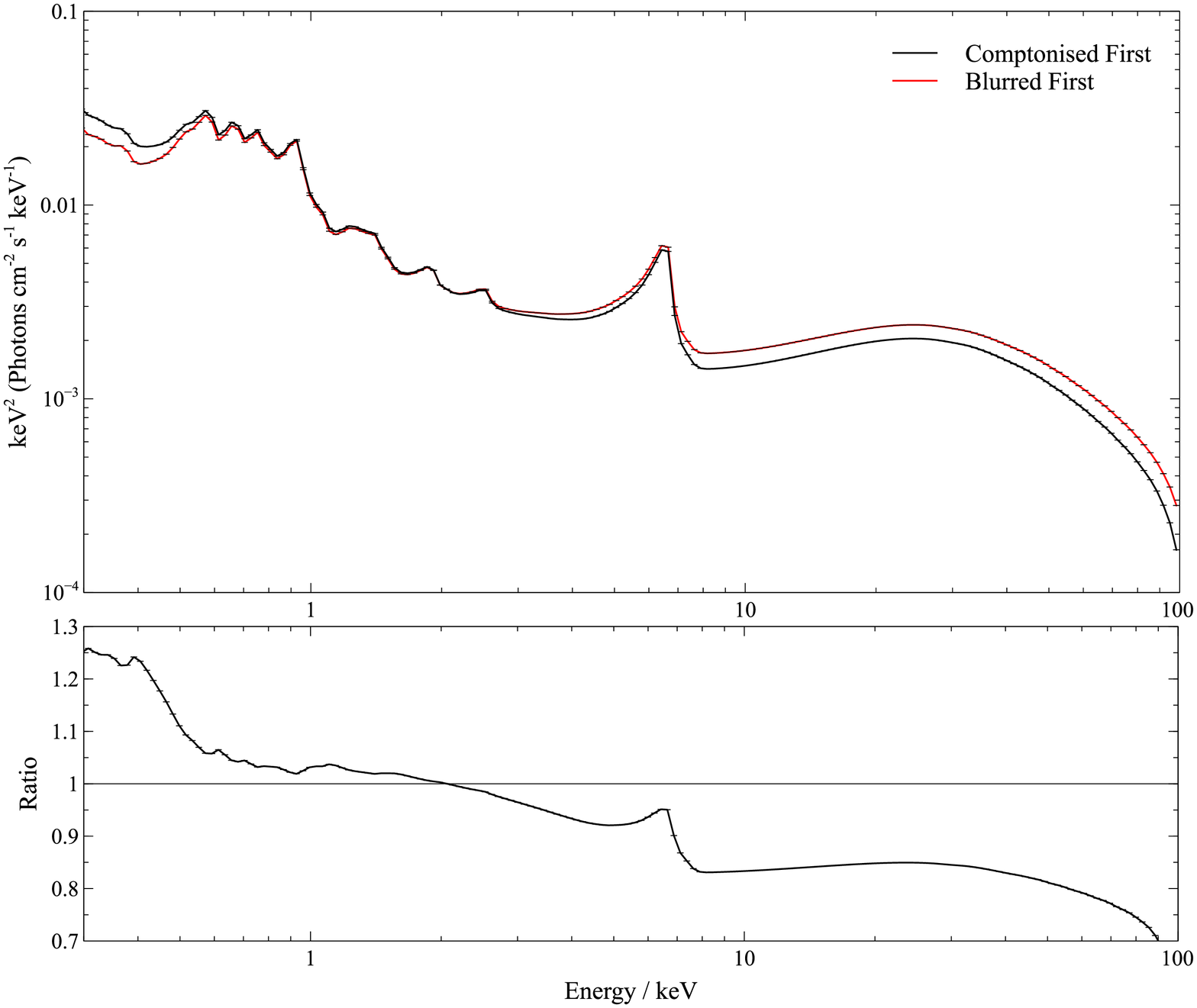}
\caption[]{Comparison of model X-ray spectra in which the Comptonisation by the corona is applied before the relativistic blurring (in the case of a corona upon the surface of the accretion disc) and in which the Comptonisation is applied to the blurred spectrum (if there is some separation between the disc and corona as assumed in the initial set of experiments). Because the features such as the iron K$\alpha$ line are sharper in the unblurred reflection spectrum, a stronger feature remains after Comptonisation when it is applied before relativistic blurring. Differences between the two cases, however, are slight and within the uncertainty of current X-ray spectroscopy possible using \textit{XMM-Newton}.}
\label{before_after_spectrum.fig}
\end{figure}

By applying the Comptonisation model to the reflection spectrum after the relativistic blurring, it is implicitly assumed that there is sufficient separation between the Comptonising medium and the accretion disc that photons reflected from any given part of the disc can be scattered by an extended section of the corona and that no energy shift is imparted on the scattered photons other than that due to Compton scattering. This is one extreme case. The opposite extreme would be a corona essentially upon and co-rotating with the surface of the accretion disc such that the unblurred reflection spectrum is scattered before the Comptonised spectrum is blurred by Doppler shifts and gravitational redshifts. In order to understand the differences between these two scenarios, the experiment was repeated interchanging the order of the \textsc{comptonise} and \textsc{kdblur3} model components (Fig.~\ref{before_after_spectrum.fig}).

The difference to the resulting spectrum is subtle and no change is found, within the errors, to the best-fitting parameters of the simple blurred reflection model between these two simulated spectra. The Comptonisation required to produce the observed X-ray continua is sufficiently strong that once Comptonised, the reflection spectrum becomes little more than a power law with no strong features, thus either the blurred reflection spectrum is smoothed to this power law or the smoothed, almost power law, spectrum is blurred which in practice, makes little difference to its shape.

These findings add credence to models such as that of \citet{merloni_fabian}, in which the X-ray emitting corona consists of a number of discrete flares, where particles (most likely electrons) above the accretion disc are accelerated to high energies by the reconnection of magnetic field lines rising from the disc. Moreover, simultaneous measurements of the ultraviolet emission from AGN, most likely in the form of thermal emission from the accretion disc, and X-ray emission rule out slab-like coron\ae\ completely covering the accretion disc in favour of patchy coron\ae\ \citep{haardt+94,dimatteo+99}.

The inferred covering fractions limit the allowed combinations of flare frequency and plasma density that will dictate how long each event lasts. It is possible that either electrons are accelerated out of the ionised disc by the flare or that there is a persistent population of electrons suspended above the disc in the magnetic field. Given the non-detection of a high-energy cut off in the continuum spectrum by hard X-ray detectors in many cases, coronal temperatures are likely in excess of 100\keV, if not 200\keV\ \citep{burlon+2011,vasudevan+13}. Therefore only optically thin populations of electrons ($\tau_\mathrm{e} < 0.5$) are required prior to their heating by magnetic reconnection events to produce the observed continuum. In regions with no flare, the reflection spectrum will, hence, not be affected by Compton up-scattering. While at any given moment, the corona is patchy, consisting of a number of isolated flares, long exposure time-averaged spectra will indicate a continuous, extended corona, through which a fraction of the reflection from the accretion disc below will be able to be seen.

Furthermore, the Comptonisation of reflected photons from the inner regions of the accretion disc could provide an explanation for an apparently truncated accretion disc, or cases where little or no relativistically blurred reflection is detected. For example, \citet{walton_6814} find that spectra of the unobscured AGN NGC\,6814 can be modelled as consisting of only a power law continuum and unblurred reflection from distant material. Na\"ively, one might expect to see relativistically blurred reflection in unobscured AGN from the accretion flow that is believed to be powering the continuum. We show here that one possible explanation for this non-detection of reflection from an accretion disc is that a dense corona of energetic particles is covering a large fraction of the inner accretion disc. It might also be that a dense corona of energetic particles covering the inner part of the accretion disc leads to an apparent truncation in the accretion disc in the canonical low hard state of some X-ray binaries and could explain why, in some sources, the disc appears to be truncated in the low hard state and in others it does not. The apparent dichotomy in which the physical conditions of this state are different between sources could be replaced by a single, continuously variable parameter between sources; the covering fraction of the inner accretion disc by the corona.

\section{Conclusions}
We find that detection of relativistically blurred reflection from the accretion discs around the supermassive black holes in AGN, right down to the innermost stable circular orbits of maximally spinning black holes at a radius of 1.235\rg\ can be self-consistently described in terms of the illuminating X-ray continuum originating from an extended corona that covers the inner part of the accretion disc, extending to tens of gravitational radii. Such a corona, however, is required to be patchy, covering a limited fraction of the surface of the disc at any one time, for instance if the corona were to consist of a number of isolated flares.

We find that it is still possible to detect reflection from the innermost stable circular orbit when the covering fraction by the patchy corona is as high as 85 per cent. Such large fractions of the reflected X-rays being Comptonised mean that the reflection fraction (the reflected flux as a fraction of the continuum flux) is systematically underestimated when not accounting for the Comptonisation of the reflected photons in spectral models. Being able to measure the high reflection fractions, $R\sim 2$ from these extended coron\ae\ requires that the covering fraction at any given moment be as low as 25 per cent.

We show that the Comptonisation of the photons reflected from the inner regions of the accretion disc can explain low reflection fractions. Reflection fractions less than unity are not expected physically when an accretion disc is illuminated by an X-ray source above the plane of the disc. High covering fractions of the inner disc by the corona cause the accretion disc to appear truncated, with a systematic overestimate in the inner radius of the disc. Measurements of other parameters of the reflection spectrum, however, including the inclination, iron abundance and ionisation, are not strongly affected by this process.

While it has been demonstrated extended coron\ae\ that cover the accretion disc can be reconciled with the detection of relativistically blurred reflection from the innermost parts of the disc beneath this corona, this does not constitute a detection of this process. We have demonstrated that up to the most extreme covering fractions, the effect on the measured spectrum is slight with the Comptonised part of the reflection spectrum blending into the primary X-ray continuum and there being only subtle changes to the broad emission lines. These results suggest, however, that the detection of low reflection fractions may point toward sources in which this process is occurring. These sources would be candidates for testing the spectral models described here and will be considered in future work. The high spectral resolution promised by X-ray microcalorimeter spectrometers, including the SXS on board Astro-H \citep{astroh} and the XIFU on Athena \citep{athena_xifu} could well prove pivotal in such studies, looking for slight decrements in the redshifted wings of broad emission lines with respect to the line peaks, where a fraction of photons at these energies are Comptonised.

\section*{Acknowledgements}
DRW is supported by a CITA National Fellowship. We thank the anonymous referee for encouraging and useful feedback that will enable us to further pursue this work.

\bibliographystyle{mnras}
\bibliography{agn}

\label{lastpage}

\end{document}